\documentclass[aps,pra,twocolumn]{revtex4}%
\usepackage{amssymb}
\usepackage{graphicx}
\usepackage{amsmath}
\usepackage{amsbsy}
\usepackage{cancel}
\usepackage{enumerate}
\usepackage{amsfonts}
\usepackage{amssymb,dsfont,physics}
\usepackage{tikz}
\usepackage{appendix}
\usetikzlibrary{decorations,snakes}
\usepackage{accents}%
\usepackage{multirow}
\usepackage{booktabs}
\providecommand{\U}[1]{\protect\rule{.1in}{.1in}}
\AtBeginDocument{
\heavyrulewidth=.08em
\lightrulewidth=.05em
\cmidrulewidth=.03em
\belowrulesep=.65ex
\belowbottomsep=0pt
\aboverulesep=.4ex
\abovetopsep=0pt
\cmidrulesep=\doublerulesep
\cmidrulekern=.5em
\defaultaddspace=.5em
}
\begin{document}
\title{Noiseless linear amplification in quantum target detection using Gaussian states}
\author{Athena Karsa}
\affiliation{Department of Computer Science, University of York, York YO10 5GH, UK}
\author{Masoud Ghalaii}
\affiliation{Department of Computer Science, University of York, York YO10 5GH, UK}
\author{Stefano Pirandola}
\affiliation{Department of Computer Science, University of York, York YO10 5GH, UK}
\date{\today}

\begin{abstract}
Quantum target detection aims to utilise quantum technologies to achieve performances in target detection not possible through purely classical means. Quantum illumination is an example of this, based on signal-idler entanglement, promising a potential 6 dB advantage in error exponent over its optimal classical counterpart. So far, receiver designs achieving this optimal reception remain elusive with many proposals based on Gaussian processes appearing unable to utilise quantum information contained within Gaussian state sources. This paper considers the employment of a noiseless linear amplifier at the detection stage of a quantum illumination-based quantum target detection protocol. Such a non-Gaussian amplifier offers a means of probabilistically amplifying an incoming signal without the addition of noise. Considering symmetric hypothesis testing, the quantum Chernoff bound is derived and limits on detection error probability is analysed for both the two-mode squeezed vacuum state and the coherent state classical benchmark. Our findings show that in such a scheme the potential quantum advantage is amplified even in regimes where quantum illumination alone offers no advantage, thereby extending its potential use. The same cannot be said for coherent states, whose performances are generally bounded by that without amplification.
\end{abstract}

\maketitle

\section{Introduction}

Quantum mechanics, and the non-classical phenomena arising from it, have revolutionised many modern technologies including computation~\cite{divincenzo1995quantum,raussendorf2001one,briegel2009measurement}, communication~\cite{gisin2007quantum,hillery1999quantum,RevQKD} and sensing~\cite{pirandola2018advances,polino2020photonic,genovese2016real,moreau2019imaging,giovannetti2011advances,braun2018quantum,pezze2018quantum}. Quantum target detection forms a particular subset of quantum sensing protocols in which ones aim is to determine whether or not a target is present in some region of interest. Quantifying one's capability of doing so, and also confirming the benefits of using a quantum strategy, is carried out on the analysis of bounds on the probability of an error, in particular, comparing the upper bound to the lower bound of the corresponding, optimal classical method. Typically this classical benchmark will take the form of a coherent state, a quantum state with minimum uncertainty, with homodyne detection at the receiver.

Quantum illumination (QI)~\cite{lloyd2008enhanced,tan2008quantum,karsa2020gensource} is one of the first proposed protocols for quantum target detection. The protocol begins by generating an entangled source comprising two modes where one is designated the role of `signal', and sent to probe the target region, while the other takes the role of the `idler' and is retained for later joint-measurement at the receiver. Remarkably, QI offers a quantum advantage in target detection despite the fact that decoherence of entanglement is encoded into the protocol itself. This quantum advantage is maximal under constraints of low signal-brightness, low target reflectivity and high background noise. Within such a regime the effective signal-to-noise ratio (SNR) of such an entangled-source transmitter offers a factor of 4 advantage over that of the corresponding classical benchmark of coherent states with homodyne detection, equivalent to a 6 dB improvement in error exponent.

Attainment of this well-known 6 dB quantum advantage through QI relies on the use of an optimal joint-measurement, however, the details of such a measurement remains unknown. Various receiver designs have been proposed for QI: the phase-conjugating (PC) and optical parametric amplification (OPA)~\cite{guha2009gaussian} achieve, at most, a 3 dB performance enhancement over coherent states while a receiver based on sum-frequency generation with feed forward (FF-SFG)~\cite{FFSFG} is capable of saturating the quantum Chernoff bound (QCB) for QI, though this receiver remains technologically out of reach.

%%% Experimentally, receivers are generally based on homodyne-type measurements carried out on the modes to determine the state's quadrature values. Zhang \textit{et. al.} \cite{zhangent}, and later \cite{zhangexp}, implemented the Gaussian QI protocol of Tan \textit{et. al}~using an OPA receiver.  

The first QI experiment by Lopaeva \textit{et al.}~\cite{lopaevaexp} used an SPDC source and photon counting to successfully demonstrate a QI-like advantage in effective SNR. More generally, receivers are based on homodyne-type measurements carried out on the modes to determine the state's quadrature values, forming better measurements than photon counting. Zhang \textit{et al.} \cite{zhangent}, and later \cite{zhangexp}, implemented the Gaussian QI protocol using an OPA receiver. Their experiment demonstrated a sub-optimal 20\% improvement (equivalent to 0.8 dB in comparison to the 3 dB available with OPA receivers) in effective SNR relative to the optimal classical scheme. There have been several initial microwave QI experiments \cite{chang2019quantum,luong2019receiver,shabirQI}. All three employ a Josephson parametric converter (JPC) for entanglement generation of microwave modes with low-brightness and compared the performance to a classically-correlated radar. The classical outcomes of heterodyne detections on each of the modes were compared in post-processing. In all experiments, a QI-like advantage was displayed over their chosen classical comparison cases. Additionally, the Barzanjeh \textit{et al.}~\cite{shabirQI} experiment compared to a coherent state source subject to the same heterodyne and post-processing receiver utilised for their entangled source.

Owing to the uncertainty principle, such homodyne-type measurements necessarily introduce noise to the system. Further, a state's homodyne statistics are described by the marginals of its Wigner function which are classical probability distributions. As such, any homodyne-type measurement on a Gaussian state, whose Wigner function is positive, results in a description of quadratures which is realistic, i.e., not purely quantum-mechanical, and thus unable to demonstrate any violation of Bell inequalities. Nonetheless, Gaussianity offers straightforward means of experimental implementation, with tools associated with Gaussian state generation, transformation and detection readily available in optics labs. As such, one could consider as an alternative either using non-Gaussian measurements on Gaussian states or Gaussian measurements on non-Gaussian states. 

One of the proposed solutions to fight loss in communication links is to use amplifiers. While standard, Gaussian amplifiers can effectively recover losses in a classical signal, they necessarily add noise to the system rendering the resultant effective SNR bounded by the original such that no overall gains in performance can be achieved. Noiseless linear amplifiers (NLAs) offer a non-Gaussian means of non-deterministically amplifying a quantum state without the addition of noise, at the expense that when the procedure fails the signal is projected onto the vacuum state and completely lost~\cite{Caves:NLA1,Caves:NLA2,Pandey:NLA,Ralph:NLA} (interested readers are referred to  Ref.~\cite{Barbieri:ReviewNLA} for a review). Experimentally, different NLA modules have been realized successfully~\cite{Ferreyrol:ExpNLA,Donaldson:ExpNLA,Chrzanowski:ExpNLA}. Previously, NLAs have been shown to demonstrate an increased robustness against loss and noise in continuous-variable quantum key distribution~\cite{Blandino:CVQKDNLA,Ghalaii:CVQKDNLA1,Ghalaii:CVQKDNLA2,Xu:CVQKDNLA} and quantum repeater~\cite{Dias:QRNLA,Ghalaii:QRNLA,Seshadreesan:QRNLA} protocols allowing for an increase in maximum transmission distance. They have also been shown to improve the performance of quantum distillation protocols~\cite{Xiang:NLAforQDist,Seshadreesan:NLAforQDist} and quantum enhancement of signal-to-noise ratio~\cite{Zhao:NLAsnu}.

In this paper we consider the use of an NLA at the detection stage of the QI protocol, effectively creating a non-Gaussian receiver, which naturally post-selects signals, for QI with a Gaussian probe. Then, by mapping the protocol of QI with a two-mode squeezed vacuum (TMSV) state with an NLA to one without an NLA but transformed Gaussian state input and quantum channel parameters, we compute the QCB. Considering the same procedure for the classical benchmark of coherent states, we show that under appropriate parameter constraints, an enhanced quantum advantage may be achieved. In particular, the NLA acting on the received TMSV quantum channel output always yields an enhancement in detection capabilities, even establishing new quantum advantages which previously did not exist. On the other hand, the resultant performance of a post-quantum channel NLA on a coherent state is, based on our analysis, always upper-bounded by the performance of a coherent state without the NLA.

\section{Noiseless linear amplification for QI}

\subsection{The QI protocol}

Consider the production of $M$ independent signal-idler mode pairs, $\{\hat
{a}_{S}^{(k)},\hat{a}_{I}^{(k)}\}$; $1\leq k\leq M$, with mean number of
photons per mode $N_{S}$ for each of the signal and idler modes,
respectively. The signal ($S$) mode is sent out to some target region while
the idler ($I$) mode is retained at the source for later joint measurement.
Their joint state, $\hat{\rho}_{S,I}$, is modelled as a two-mode, zero-mean
Gaussian state~\cite{RMP} with covariance matrix (CM) given by
\begin{equation}
\mathbf{V}_{S,I}=
\begin{pmatrix}
\nu\mathbf{1} & c_q\mathbf{Z}\\
c_q\mathbf{Z} & \nu\mathbf{1}%
\end{pmatrix}
,~\left\{
\begin{array}
[c]{l}%
\mathbf{1}:=\mathrm{diag}(1,1),\\
\mathbf{Z}:=\mathrm{diag}(1,-1),
\end{array}
\right.  \label{eq1}%
\end{equation}
where $\nu:=2N_{S}+1$ and $c_q=2\sqrt{N_{S}(N_{S}+1)}$ quantifies the quadrature
correlations between the two modes. The off-diagonal terms can in fact take any value such that $0\leq c\leq2\sqrt{N_{S}%
(N_{S}+1)}$. In the case where the signal-idler mode pairs are maximally
entangled we have $c=c_{q}:=2\sqrt{N_{S}(N_{S}+1)}$ (the TMSV state~\cite{RMP}) while the case
$c=c_{d}:=2N_{S}$ renders the state
just-separable~\cite{EntBreak,ModiDiscord}.

Under hypothesis $H_{0}$, the target is absent so that the returning mode
$\hat{a}_{R}=\hat{a}_{B}$, where $\hat{a}_{B}$ is in a thermal state with mean
number of photons per mode $N_{B}$. Under hypothesis $H_{1}$, the target
is present such that $\hat{a}_{R}=\sqrt{\kappa}\hat{a}_{S}+\sqrt{1-\kappa}%
\hat{a}_{B}$. Here, $\kappa$ is the target reflectivity, incorporating all propagation losses associated with the channel, and $\hat{a}_{B}$ is in a thermal state with
mean number of photons per mode $N_{B}/(1-\kappa)$, so that the mean noise
photon number is equal under both hypotheses (i.e., there is no passive signature and a non-vacuum transmitter must be used in order to detect the target). The conditional joint state,
$\hat{\rho}_{R,I}^{i}$ for $i=0,1$, of the returning ($R$) mode and the
retained idler ($I$) is given by, under hypotheses $H_{0}$ and $H_{1}$,
respectively,
\begin{equation}
\mathbf{V}_{R,I}^{0}=
\begin{pmatrix}
\omega\mathbf{1} & 0\\
0 & \nu\mathbf{1}%
\end{pmatrix}
, \label{eq2}%
\end{equation}%
\begin{equation}
\mathbf{V}_{R,I}^{1}=
\begin{pmatrix}
\gamma\mathbf{1} & \sqrt{\kappa}c_q\mathbf{Z}\\
\sqrt{\kappa}c_q\mathbf{Z} & \nu\mathbf{1}%
\end{pmatrix}
, \label{eq3}%
\end{equation}
where we set $\omega:=2N_{B}+1$ and $\gamma:=2\kappa N_{S}+\omega$.

\begin{figure}[t!]
    \centering
\begin{tikzpicture}
\draw[-,red!40,ultra thick] (-0.5,0.5) -- (6.2,0.5) -- (6.2,-4.6) -- (-0.5,-4.6) -- (-0.5,0.5);
\node at (0.3,-1) {\includegraphics[width=0.055\textwidth]{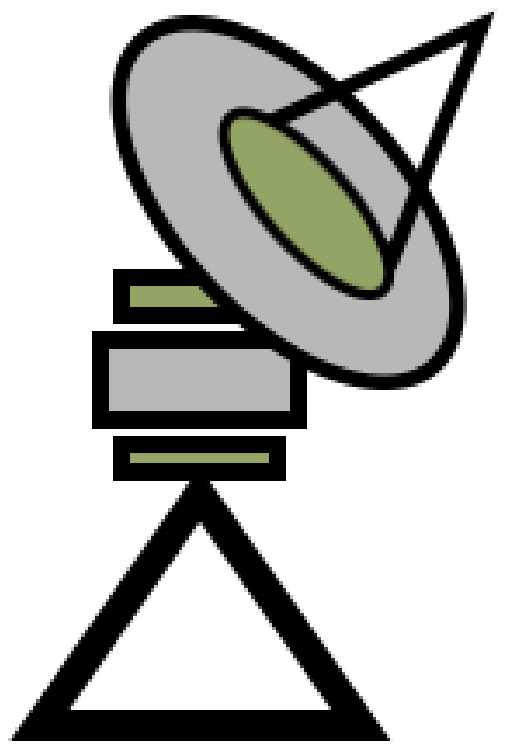}};
\draw[->, black, thick] (1,-0.5) --(2.75,-0.5);
\draw[->, black, thick] (2.75,-0.5) --(5,-0.5);
\node at (2.75,-0.1) {$\hat{a}_S^{(k)}$};
\node at (4,-2) {$\hat{a}_R^{(k)}$};
\draw[-, black,ultra thick] (4.5,-0.1) --(5.5,-0.9);
\node at (5.4,-0.2) {target};
\node at (5.5,-1.1) {$\kappa$};
\draw[->, black, thick] (5,-0.5) -- (2.5,-3.1);
\draw[->,red!40, snake=snake, thick] (4.8,-3.55)-- (3,-3.55);
\node at (5.4, -3.35) {$\rho_{th}(N_B)$};
\node at (5.4, -3.75) {$\hat{a}_B$};
\node at (3.9,-3.25) {$(1-\kappa)$};
\draw[-,black!60,ultra thick] (2,-3.2) -- (2,-3.8)--(3,-3.8)-- (3,-3.2)--(2,-3.2);
\node at (2.5,-3.5) {NLA};
\draw [->,black,thick] (1,-0.5)--(1,-3.45);
\draw [->,black,thick] (1.95,-3.55)--(0.95,-3.55);
\node at (0.8,-2.2) {$\hat{a}_I^{(k)}$};
\begin{scope}
\clip (0.8,-4) rectangle (0.3,-3);
\draw[fill=black!45] (0.8,-3.5) circle(0.4);
\end{scope}
\draw[-,black!45, snake=snake, ultra thick] (0.6,-3.5)-- (-0.4,-3.5);
\end{tikzpicture}
    \caption{Protocol for QI with the use of NLA at the detector. $M$ independent signal-idler source mode pairs are generated with annihilation operators $\hat{a}_S^{(k)}$ and $\hat{a}_I^{(k)}$, respectively, with $1 \leq k \leq M$. The signal mode is sent to probe the target region in which target of reflectivity $\kappa$ is equally-likely to be present or absent while the idler mode is sent straight to the receiver. At the receiver, the returning signal, mixed with the ambient background $\hat{a}_B$, first encounters an NLA which probabilistically noiselessly amplifies it before recombination with the idler in the decision-making process.}
    \label{fig:roomCS}
\end{figure}

\subsection{NLA action and effective parameters for QI}

Consider the entanglement-based QI protocol where the source is a TMSV state comprising signal and idler modes given by
\begin{equation}
    \ket{\lambda}_{S,I} = \sqrt{1-\lambda^2} \sum_{n=0}^{\infty} \lambda^n
\ket{n}_S\ket{n}_I,
\end{equation}
with $\lambda^2=\frac{N_S}{N_S+1} <1$, where $N_S$ is the average number of photons per mode. Its initial CM is equivalent to that in Eq.~(\ref{eq1}).

Consider the action of a generic Gaussian channel with transmissivity $\tau$, and excess noise $\epsilon$ on a single mode $A$ of an arbitrary input TMSV state with CM $\gamma_{A,B}$. The output CM is given by
\begin{equation}\label{candCM}
    \gamma_{A,B}' = 
\begin{pmatrix}
\tau(V +B+\epsilon)\mathbf{1} & \sqrt{\tau(V^2-1)}\mathbf{Z}\\
\sqrt{\tau(V^2-1)} \mathbf{Z} & V\mathbf{1}%
\end{pmatrix},
\end{equation}
where $V=(1+\lambda^2)/(1-\lambda^2)$ is the variance of the thermal state $\Tr_A \dyad{\lambda}{\lambda}$ and $B=(1-\tau)/\tau$ is the input equivalent noise due to losses.

Now consider the implementation of a NLA to mode $A$ prior to measurement. It can be shown that (see Ref.~\cite{Blandino:CVQKDNLA}, particularly App. A) the CM $\gamma_{A,B}'(\lambda, \tau, \epsilon, g)$ of the amplified state, post NLA action, is equivalent to the CM $\gamma_{A,B}'(\lambda^g, \tau^g, \epsilon^g, g=1)$ of an equivalent system with TMSV parameter $\lambda^g$, under action of a Gaussian channel with transmissivity $\tau^g$ and excess noise $\epsilon^g$, without the use of an NLA. These effective parameters are given by
\begin{align}\label{eq:param}
\lambda^g = & \lambda \sqrt{\frac{(g^2-1)(\epsilon - 2)\tau-2}{(g^2-1)\epsilon \tau -2}}, \notag \\
\tau^g = & \frac{g^2 \tau}{(g^2 -1)\tau \big(\frac{1}{4}(g^2-1)(\epsilon-2) \epsilon \tau - \epsilon +1\big) +1}, \notag \\
\epsilon^g = & \epsilon - \frac{1}{2}(g^2-1)(\epsilon-2) \epsilon \tau. 
\end{align}
For the above system of effective parameters to represent an actual physical system, the following constraints must be satisfied: $0 \leq \lambda^g < 1$, $0 \leq \tau^g \leq 1$ and $\epsilon^g \geq 0$. The first is always satisfied when
\begin{equation}\label{eq:constraint1}
    0 \leq \lambda^g < 1 \Rightarrow 0 < \lambda < \left(\sqrt{\frac{(g^2-1)(\epsilon -2)\tau -2}{(g^2-1)\epsilon \tau -2}} \right)^{-1}.
\end{equation}
The second and third conditions are satisfied provided the excess noise $\epsilon <2$ and the gain is smaller than a maximum value given by
\begin{equation}\label{conditiong}
\begin{split}
    &g_{\mathrm{max}}=\\
    &\resizebox{.9\hsize}{!}{$\sqrt{\frac{\epsilon (\tau(\epsilon-4)+2)+4\sqrt{\frac{\tau(\epsilon-2)+2}{\epsilon}} -2\sqrt{\epsilon(\tau(\epsilon-2)+2)} +4\tau -4}{\tau(\epsilon-2)^2}}$}.
\end{split}
\end{equation}

Equivalences can be made between Eq.~(\ref{candCM}) and Eq.~(\ref{eq3}): For QI we consider a TMSV state with $N_S$ mean photons per mode such that the variance $V=2N_S+1$ and $\sqrt{V^2-1} = 2\sqrt{N_S(N_S+1)}$ while Gaussian channel transmissivity $\tau \equiv \kappa$, the target reflectivity. Of course, for real-world target detection this parameter would also incorporate other losses and gains given by the radar equation. In QI, a portion $\kappa$ of the signal is mixed with the thermal background, which comprises $N_B/(1-\kappa)$ mean photons per mode. Taking into account this rescaling, when the target is present the returning signal mode takes the form
\begin{equation}
\begin{split}
    &\kappa (2 N_S + 1) + (1-\kappa) \left( \frac{2 N_B}{1-\kappa}+1\right)\\
    &\equiv \tau V + B \tau \left( \frac{2 N_B}{1-\tau}+1\right)\\
    &= \tau\left( V + B  + \frac{2 N_B}{\tau}\right) \equiv \tau(V+B+\epsilon),
\end{split}
\end{equation}
where excess noise has a simple relation with $N_B$ given by $\epsilon = \frac{2N_B}{\tau} \equiv \frac{2N_B}{\kappa}$. Thus by considering an equivalent system of effective parameters in place of the two conditional CMs for QI given in Eqs.~(\ref{eq2}) and~(\ref{eq3}), one can consider the additional action of an NLA on the returning signal modes at the receiver, before joint measurement with the retained idler.

Note that the constraint on excess noise to maintain the effective system's physicality means that $\epsilon = \frac{2 N_B}{\kappa} < 2 $, i.e., $N_B<\kappa$. Since $0\leq \kappa\leq 1$, we have the global constraint $N_B<1$ on the mean number of thermal photons associated with the background. Typically, for QI, the parameter constraints involve very high background, $N_B\gg1$, which is naturally satisfied in the microwave domain at room temperature, and $\kappa \ll 1$. However these are not strictly necessary for a quantum advantage exists; provided $N_S \ll 1$ quadrature correlations $c_q$ are maximised and it is from here where the quantum advantage arises. The new constraint on $N_B$ introduced here means that, comfortably, at room temperature ($T=300$K) applications the protocol described here is valid for frequencies $\gtrsim 4$THz, beginning at the higher end of the microwave. Lower frequencies can meet this requirement as long as the temperature of application is small enough, e.g., for operations at $\sim 1$GHz we require $T\lesssim 0.07$K.

Further, for a given environment ($N_B$) and target parameters ($\kappa$), Eq.~(\ref{eq:constraint1}) implies that the maximum value of signal energy, $N_S$, which may be employed is given by
\begin{equation}
    N_S^{\mathrm{max}}(g) = \frac{1-N_B(g^2-1)}{\kappa (g^2-1)},
\end{equation}
which is maximised when $g=1$, i.e., no amplification occurs and the protocol is equivalent to that of standard QI.

The action of the NLA is a non-deterministic one. That is, it provides a tool for heralded noiseless quantum amplification, i.e., ideally, the transformation $\ket{\alpha}\rightarrow\ket{g \alpha}$, where $g>1$ is the NLA gain, with some probability of success, $P(g)$ \cite{Caves:NLA2,Pandey:NLA,Ralph:NLA}. 
%bounded by
%\begin{equation}\label{eq:psucc}
%    P_{\mathrm{succ}}^{\mathrm{NLA}} = \frac{1}{g^2},
%\end{equation}
In other words, under NLA action the number of probings used for the detection process transforms as $M \rightarrow M P(g)$ with the remaining $M(1-P(g))$ channel uses discarded. 
Thus, with NLA action we are considering post-selected QI and the problem of hypothesis testing becomes one of two stages and four potential outcomes:

\begin{description}
\item[$H_{00}:$] Target is absent, and the NLA is unsuccessful;
\item[$H_{01}:$] Target is absent, and the NLA is successful;
\item[$H_{10}:$] Target is present, and the NLA is unsuccessful;
\item[$H_{11}:$] Target is present, and the NLA is successful.
\end{description}
Post-selection essentially discards all events corresponding to hypotheses $H_{00}$ and $H_{10}$ and the problem is reduced to standard QI involving the discrimination of only two hypotheses $H_{01}$ and $H_{11}$, subject to $M \rightarrow MP(g)$.

Note that on average, taking into account all successful and unsuccessful NLA outcomes, the distinguishability of the quantum states does not increase. However, a scheme where successful amplifications are heralded (see Ref.~\cite{Ralph:NLA}) such that measurements are only performed on successfully amplified outputs can yield performance enhancements in various protocols~\cite{nlaFisherInformation}.

\subsection{Classical benchmarking with coherent states}

In the absence of an idler, and for the purposes of defining a classical benchmark for our approach, we consider the coherent state as the optimal corresponding classical approach. Coherent states are minimum-uncertainty quantum states which may be employed in analogous protocols but whose statistics originate from purely classical phenomena. As such, they can serve as theoretically optimal classical states to benchmark potential quantum protocols against, thus allowing one to define a quantum advantage.

The signal is prepared in the coherent state $|\sqrt{2N_{S}}\rangle$ which is then sent out to some target region. Under $H_{0}$, the received returning mode is in a thermal state with mean photon number $N_{B}$ and CM equal to $\omega\mathbf{1}$, i.e., $\hat{a}_{R}=\hat{a}_{B}$. Under $H_{1}$, the signal is mixed with the background such that $\hat{a}_{R}=\sqrt{\kappa}%
\hat{a}_{S}+\sqrt{1-\kappa}\hat{a}_{B}$ with $\kappa\in(0,1)$, corresponding to a displaced thermal state with mean vector $(\sqrt{2\kappa N_{S}},0)$ and CM $\omega \mathbf{1}$. 

Consider the thermal state $\hat{\rho}_{\mathrm{th}}(\lambda_{\mathrm{th}})$ with Fock basis representation
\begin{equation}
    \hat{\rho}_{\mathrm{th}}(\lambda_{\mathrm{th}}) = (1-\lambda_{\mathrm{th}}^2) \sum_{n=0}^{\infty} \lambda_{\mathrm{th}}^{2n} \dyad{n}{n},
\end{equation}
displaced by complex $\beta$ yielding the state $\hat{\rho} = \hat{D}(\beta) \hat{\rho}_{\mathrm{th}}(\lambda_{\mathrm{th}}) \hat{D}(-\beta)$. Such a state can be written as an ensemble of coherent states,
\begin{equation}
    \hat{\rho} = \int P(\alpha) \dyad{\alpha}{\alpha} \mathrm{d}\alpha,
\end{equation}
where $P(\alpha)=\frac{e^{|\alpha|^2}}{\pi^2} \int e^{|u|^2} \langle -u | \hat{\rho} | u \rangle e^{u^{\star}\alpha - u \alpha ^{\star}} \mathrm{d}u$, is the $P$-function~\cite{Blandino:CVQKDNLA}.

After successful amplification, realised by the operator $\hat{C}=g^{\hat{n}}$ where $\hat{n}$ is the Fock basis number operator, the coherent state $\ket{\alpha}$ transforms as
\begin{equation}
    \hat{C}\ket{\alpha} = e^{\frac{|\alpha|^2}{2}(g^2 -1) \ket{g \alpha}}
\end{equation}
such that the initial state after NLA action becomes
\begin{equation}
    \hat{\rho}' = \hat{C}\hat{\rho}\hat{C} = \int P(\alpha) e^{|\alpha|^2(g^2-1)} \dyad{g \alpha}{g \alpha} \mathrm{d}\alpha.
\end{equation}
After change of variables it can be found that the resulting state after NLA action obeys the following relation of proportionality:
\begin{equation}
    \hat{\rho}' \propto \hat{D}(\bar{g}\beta)\hat{\rho}_{\mathrm{th}} (g \lambda_{\mathrm{th}}) \hat{D}(-\bar{g} \beta),
\end{equation}
where $\bar{g} = g(1-\lambda_{\mathrm{th}}^2)/(1-g^2 \lambda_{\mathrm{th}}^2)$. That is, as in the case for a TMSV source, the result of a displaced thermal state acted on by an NLA is equivalent to a displaced thermal state with modified effective parameters without amplification, subject to the constraint that $g \lambda_{\mathrm{th}}<1$ to ensure physicality.

For QI applications, the initial coherent state $\ket{\sqrt{2 N_S}}$ is sent through a quantum channel with reflectivity/transmittance $\kappa$ such that the displacement can be taken as $\beta = \sqrt{2 \kappa N_S}$. Meanwhile, the variance of the thermal state is given by
\begin{equation}
    \frac{1+ \lambda_{\mathrm{th}}^2}{1-\lambda_{\mathrm{th}}^2} = 2 N_B + 1 =\omega \Rightarrow \lambda_{\mathrm{th}}^2 = \frac{N_B}{1+N_B}.
\end{equation}
Thus, action of the NLA on the displaced thermal state with these parameters yields the following transformations: for the mean,
\begin{equation}\label{CSeffectiveMEAN}
\begin{split}
    \sqrt{2 \kappa N_S} &\rightarrow g\frac{1-\lambda_{\mathrm{th}}^2}{1-g^2 \lambda_{\mathrm{th}}^2} \sqrt{2 \kappa N_S} \\
    & = \frac{g}{1+N_B (1-g^2)} \sqrt{2\kappa N_S} = \beta'.
\end{split}
\end{equation}
Then,
\begin{equation}
   \lambda_{\mathrm{th}}^2 = \frac{N_B}{1+N_B} \rightarrow g^2 \frac{N_B}{1+N_B} = \lambda_{\mathrm{th}}'^2,
\end{equation}
such that the effective variance becomes
\begin{equation}\label{CSeffectiveVAR}
     \frac{1+ \lambda_{\mathrm{th}}'^2}{1-\lambda_{\mathrm{th}}'^2} = \frac{1 + N_B(1 + g^2)}{1 + N_B(1 -g^2)} = \omega'.
\end{equation}

\subsection{Performance bounds for QI with NLA}

\subsubsection{TMSV state with NLA}\label{sec:TMSVQCB}

Using the tools of App.~\ref{sec:QCBtools} the QCB of the maximally-entangled TMSV source for QI may be computed. We compute the QCB for a single successful use of the QI protocol, where the probability of success is given by that of the NLA which in the ideal case is equal to $P_{\mathrm{succ}}^{\mathrm{NLA}} = 1/g^2$, where $g$ is the NLA gain. Let us denote the QCB on the output of a successful NLA following QI as $\xi_{\mathrm{QI+NLA}}$. Then, for equally likely hypotheses, and after $k$ uses, the QCB becomes
\begin{equation}
    \frac{1}{2}\left(\xi_{\mathrm{QI+NLA}}\right)^k.
\end{equation}
But, we must include the probabilistic nature of successful outcomes. Given a total of $M$ uses of the entire QI protocol, that is $M$ copies of the TMSV source, discrimination is carried out on the $k$ successful outcomes where $k$ follows a binomial distribution. Then, the total average error probability $P_{\mathrm{QI+NLA}}^{\mathrm{QCB}}$, for equally likely hypotheses, becomes
\begin{equation}\label{binomialqcb}
\begin{split}
   & \frac{1}{2}\sum_{k=0}^M \left( \xi_{\mathrm{QI+NLA}}\right)^k {M \choose k} \left( P_{\mathrm{succ}}^{\mathrm{NLA}} \right)^k (1-P_{\mathrm{succ}}^{\mathrm{NLA}})^{M-k} \\
   & = \frac{1}{2} \left(1 + P_{\mathrm{succ}}^{\mathrm{NLA}} (-1 + \xi_{\mathrm{QI+NLA}}) \right)^M\\
   &=\frac{1}{2} \left( 1+ \frac{1}{g^2} (-1 +\xi_{\mathrm{QI+NLA}} ) \right)^M = P_{\mathrm{QI+NLA}}^{\mathrm{QCB}},
\end{split}
\end{equation}
where in the last line we have set the ideal case of $P_{\mathrm{succ}}^{\mathrm{NLA}} = 1/g^2$. Note that in the case of no successful NLA outcomes, i.e., $k=0$, one simply chooses at random, bounding the maximum error to be $1/2$, such that regardless of how the NLA behaves a decision is always made as to whether or not the target is present. Further, all measurements, and thus the entirety of decision-making, are only based on the $k$ successful NLA output states.

\begin{figure}[t]
    \centering
    \includegraphics[width=0.9\linewidth]{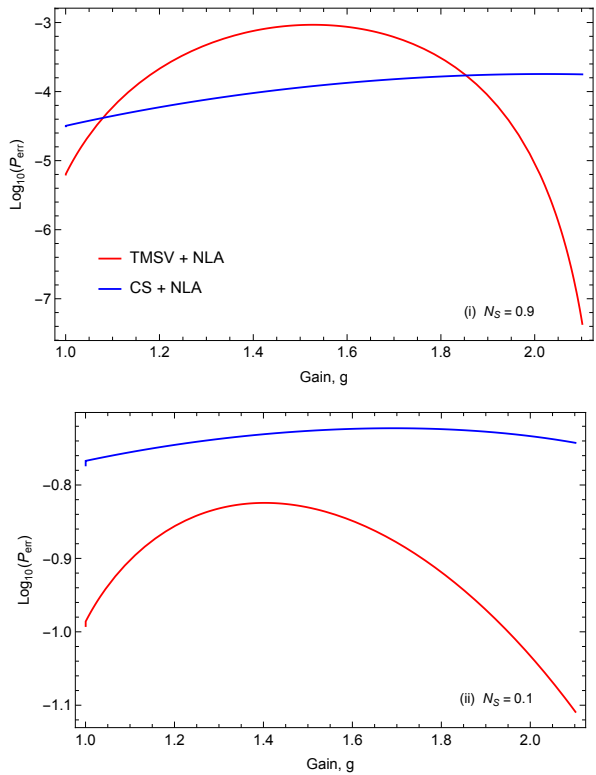}
    \caption{Error probability exponents for QI using a maximally-entangled TMSV source with NLA (red) at the receiver, compared to a coherent state source with the same NLA (blue) as a function of NLA gain, $g$. In both panels, parameters are set such that $N_B=0.1$, $\kappa=0.2$ such that the maximum source energy applicable across the range, $N_S^{\mathrm{max}}(g_{\mathrm{max}}) \simeq 0.96$. Thus, values are plotted for (i) $N_S=0.9$ and (ii) $N_S=0.1$. The total number of probes $M=100$.}
    \label{fig:NLAvsgain1}
\end{figure}

The computation is carried out using mathematical computational software and, while the full form too long to be exhibited here, its behaviour is plotted in Figs.~\ref{fig:NLAvsgain1} and~\ref{fig:NLAvsgain2} and discussed in Sec.~\ref{sec:benchmarking}.

\subsubsection{Coherent state with NLA}

As with the TMSV source, the tools of App.~\ref{sec:QCBtools} may be used to compute the QCB of a coherent state with amplification by considering an equivalent protocol, without amplification, using modified effective parameters for mean and variance given by Eqs.~(\ref{CSeffectiveMEAN}) and (\ref{CSeffectiveVAR}), respectively. For equally-likely hypotheses, the single-use ($M=1$) QCB for a coherent state with NLA amplification takes the exact form
\begin{equation}\label{csQCBNLA1}
\begin{split}
    P^{\mathrm{QCB},M=1}_{\mathrm{CS+NLA}} &= \frac{1}{2} \exp \left(- \frac{g^2 \kappa N_S \left(\sqrt{N_B+1} - g\sqrt{N_B}\right)^2}{(1 + N_B - g^2 N_B)^3}\right) \\
    &= \frac{1}{2} \xi_{\mathrm{CS+NLA}},
    \end{split}
\end{equation}
assuming successful amplification for that single use.

Taking into account the probabilistic nature of our NLA procedure, the overall average error probability for the coherent state source follows the same behaviour as seen for the TMSV state for QI with an NLA. Explicitly, after $M$ uses with $k$ successes, the average error probability becomes
\begin{equation}\label{binomialqcbCS}
    P_{\mathrm{CS+NLA}}^{\mathrm{QCB}} = \frac{1}{2} \left( 1+ \frac{1}{g^2} (-1 +\xi_{\mathrm{CS+NLA}} ) \right)^M.
\end{equation}
Meanwhile, the QCB of such a coherent state
transmitter, without amplification, may be readily computed and takes the exact
form~\cite{tan2008quantum}
\begin{equation}\label{extracsQCB}
P^{\mathrm{QCB}}_{\mathrm{CS}}=\frac{1}{2}\exp\left(-M\kappa N_{S}\left(  \sqrt{N_{B}%
+1}-\sqrt{N_{B}}\right)  ^{2}\right). %
\end{equation}

Note that the QCB forms an upper bound to the minimum error probability, i.e., $P^{\mathrm{QCB}} \leq P^{\mathrm{min}}(e)$, which is exponentially tight in the limit $M \rightarrow \infty$. To be certain of our comparisons and the determination of any quantum enhancement via NLA use, we can perform comparisons to the lower bound on the error probability for coherent states (see App.~\ref{sec:QCBtools}, Eq.~(\ref{errorLB})). Explicitly, for coherent states without an NLA, this takes the exact form
\begin{equation}
\begin{split}
    &P^{\mathrm{min}}_{\mathrm{CS}} \geq\\
    &\frac{1}{2} \left(1- \sqrt{1- \exp\left(-2M\kappa N_{S}\left(  \sqrt{N_{B}%
+1}-\sqrt{N_{B}}\right)  ^{2}\right)} \right).
\end{split}
\end{equation}

% Taking into account the number of successful probings used to achieve this bound, each occurring with probability of success $P(g)$, we have for a total of $M$-uses, with $M>1$, 
% \begin{equation}\label{csQCBNLAM}
% \begin{split}
%     &P^{\mathrm{QCB}}_{\mathrm{CS+NLA}}\leq \\
%     &\frac{1}{2} \exp \left(- \frac{M P(g) g^2 \kappa N_S \left(\sqrt{N_B+1} - g\sqrt{N_B}\right)^2}{(1 + N_B - g^2 N_B)^3}\right),
%     \end{split}
% \end{equation}
% with similar consideration employed in the computation of the TMSV + NLA QCB in Sec.~\ref{sec:TMSVQCB}.

% The QCB of such a coherent state
% transmitter, without amplification, may be readily computed and takes the exact
% form~\cite{tan2008quantum}
% \begin{equation}\label{csQCB}
% P^{\mathrm{QCB}}_{\mathrm{CS}}\leq\frac{1}{2}\exp\left(-M\kappa N_{S}\left(  \sqrt{N_{B}%
% +1}-\sqrt{N_{B}}\right)  ^{2}\right). %
% \end{equation}

With the addition of an NLA at the detection stage, it is possible to establish a gain in performance in a quantum-inspired coherent illumination protocol, but only with respect to associated upper bounds in detection error probability. When comparing to the lower bound, the limitations imposed by our choice of analysis, namely the parameter constraints outlined in Eq.~(\ref{eq:param}), render the regimes in which this is possible physically out of bounds. Thus, a quantum-inspired illumination protocol based on a coherent state source using an NLA at the receiver cannot absolutely provide a means for improved target detection. However, this does not preclude the possibility that an alternative analysis exists, which does not impose the same constraints as that studied, making such a protocol useful.

\subsection{Benchmarking QI with NLA}\label{sec:benchmarking}

\subsubsection{Comparison of NLA protocols}\label{sec:NLAcomp}

Since the coherent state forms the ideal, minimum-uncertainty state and serves as the theoretically optimal classical benchmark, Eq.~(\ref{binomialqcbCS}) allows for the benchmarking of the TMSV with the use of an NLA for target detection.

\begin{figure}[t]
    \centering
    \includegraphics[width=0.9\linewidth]{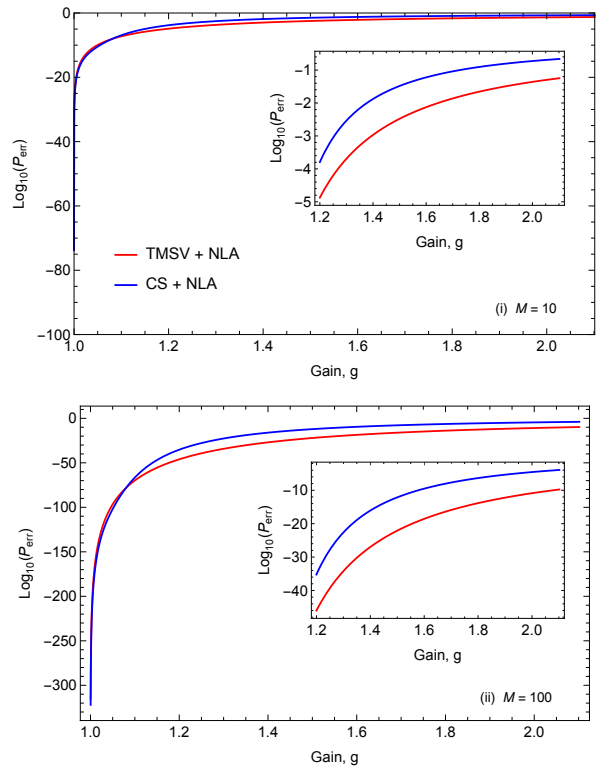}
    \caption{Error probability exponents for QI using a maximally-entangled TMSV source with NLA (red) at the receiver, compared to a coherent state source with the same NLA (blue) as a function of NLA gain, $g$. In both panels, parameters are set such that $N_B=0.1$, $\kappa=0.2$, while for each value of $g$, the signal energy is set very close (99\%) to its local maximum, i.e., $N_S = N_S^{\mathrm{max}}(g)$. Total number of probes is set to for (i) $M=10$ and (ii) $M=100$.}
    \label{fig:NLAvsgain2}
\end{figure}

Taking into account constraints on effective parameters given by Eq.~\eqref{eq:param}, Figs.~\ref{fig:NLAvsgain1} and~\ref{fig:NLAvsgain2} plot the performance of the TMSV state with NLA relative to that of a coherent state with NLA. Note that the full, exact forms of the QCB have been employed in the computation, that is, without any assumptions as to the relative magnitude of parameter values. Further, the plots have been generated assuming a maximum theoretical probability of success, given by $P_{\mathrm{succ}}^{\mathrm{NLA}} = 1/g^2$ to model the absolute limits of NLA performance.

In Fig.~\ref{fig:NLAvsgain1}, the error probability exponent is plotted as function of the NLA gain, $g$, up to and including $g_{\mathrm{max}}$, for fixed environmental parameters $N_B=0.1$ and $\kappa=0.2$ with the total number of probings $M=100$. Based on these parameters it can be found that the maximum energy valid across all values of $g$, maintaining physicality, is given by $N_S^{\mathrm{max}}(g_{\mathrm{max}}) \simeq 0.96$ thus results are plotted for two values of $N_S$: $0.9$ and $0.1$. It can clearly be seen that an increase in the gain, $g$, has a much larger and more valuable effect on the the TMSV state, compared to the same amplification of the returning coherent state. Note that where $g=1$ the performance coincides with that of the standard QI protocol without any amplification. As expected, smaller values of source energy $N_S$ are favoured by the QI with a TMSV source compared to the coherent state since it is for small $N_S$ where cross-correlations, $c_q =2\sqrt{N_S(N_S+1)}$, are maximised.

Fig.~\ref{fig:NLAvsgain2} plots the same function as Fig.~\ref{fig:NLAvsgain1} with much of the same parameters, however in this scenario rather than considering the global maximum of $N_S$, applicable across all values of $g$, up to and including $g_{\mathrm{max}}$, we consider a source whose energy is given by (99\% of) the local maximum. That is, for each value of $g \in \left[0,g_{\mathrm{max}}\right]$, $N_S$ is set such that $N_S=N_S^{\mathrm{max}}(g)$. Of course, $N_S^{\mathrm{max}}$ is a decreasing function of $g$ so the behaviour observed for $g \rightarrow 1$, where $N_S$ is typically very large, the coherent state outperforms the TMSV. However small increases in $g$ show a large quantum advantage can be achieved, even at the maximal $N_S$ value. This quantum advantage may be amplified in cases where the source energy must be kept low, as in stealth surveillance or biomedical sensing where samples may be sensitive to high energies, due to the freedom available in decreasing $N_S$ below the value used in this comparison. Making use of such freedom will, of course, amplify entanglement benefits.

\subsubsection{Comparison with non-NLA protocols}\label{sec:nonNLAcomp}

While Sec.~\ref{sec:NLAcomp} shows that the use of NLAs yields improvement in performance for TMSV protocols over coherent state protocols, there is, of course, a question as to whether or not their use is beneficial when one can simply forgo the NLA and keep all $M$ channel uses in the detection. After all, successful amplification comes at the expense of a proportion, QCBs for target detection, both for QI with a TMSV source and coherent states (see Ref.~\cite{karsa2020gensource} for full details), may be recovered by simply setting $g=1$.

\begin{figure}[t!]
    \centering
    \includegraphics[width=0.9\linewidth]{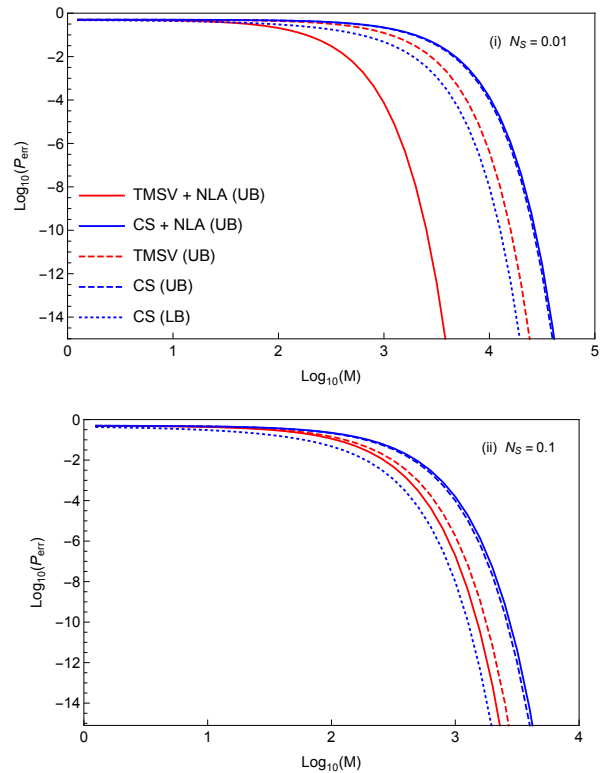}
    \caption{Error probability exponents for QI using a maximally-entangled TMSV source with NLA (red, solid) at the receiver, compared to a coherent state source with the same NLA (blue, solid) as a function of the number of probes, $M$. Also included are associated upper bounds without the use of the NLA (dashed) showing how the NLA enhances this upper bound. The lower bound for coherent states (blue, dotted) shows the point beyond which a quantum enhancement (through the NLA, entanglement, or both) may be confirmed.} In both panels, parameters are set such that $N_B=0.1$, $\kappa=0.2$ such that the maximum source energy applicable across the range, $N_S^{\mathrm{max}}(g_{\mathrm{max}}) \simeq 0.96$, with $g=g_{\mathrm{max}}\simeq 2.1$. Thus, values are plotted for (i) $N_S=0.01$ and (ii) $N_S=0.1$.
    \label{fig:NLAvsMcomp1}
\end{figure}

% \begin{figure}[t]
%     \centering
%     \includegraphics[width=0.9\linewidth]{NLAcomp_nb0002_k01new.eps}
%     \caption{Error probability exponents for QI using a maximally-entangled TMSV source with NLA (red, solid) at the receiver, compared to a coherent state source with the same NLA (blue, solid) as a function of the number of probes, $M$. Also included are performance bounds without the use of the NLA (dashed). In both panels, parameters are set such that $N_B=0.005$, $\kappa=0.1$ such that the maximum source energy applicable across the range, $N_S^{\mathrm{max}}(g_{\mathrm{max}}) \simeq 0.075$, with $g=g_{\mathrm{max}}/2\simeq 5.5$. Values are plotted for $N_S=0.07$.}
%     \label{fig:NLAvsMcomp2}
% \end{figure}

Fig.~\ref{fig:NLAvsMcomp1} plots the error probability exponents for QI using a maximally entangled TMSV source with an NLA at the receiver alongside that of a coherent state source using the same NLA. For comparison and to show that the NLA is of actual value, we plot the QCBs for the same protocols without the use of the NLA. In these protocols all $M$ probings are used at the receiver in decision-making. Results show that there exists a clear advantage in employing NLAs at the receiver compared to without.

With respect to the determination of a true quantum advantage, we also compare to the coherent state lower bound on the minimum error probability (which is not exponentially tight in the number of uses). For small signal energies, the employment of an NLA at the receiver for QI-based quantum target detection allows for the realisation of a \emph{new} quantum advantage that was previously unavailable. Within these low brightness regimes, it can be seen from the plot and numerical calculations that a factor of $\sim 5$ (equivalent to $\sim 7$dB) advantage in error exponent (based on QI's upper bound in error probability) can be established over the coherent state's lower bound by utilising the NLA. For comparison, in this regime the effective signal-to-noise ratio for QI without the NLA exhibits a $\sim 3$dB loss with respect to the coherent state lower bound.

\subsubsection{Effect of sub-optimal NLA success probability}

Hitherto the assumption has been made that the NLA in use at the detection stage of QI is ideal. Such an ideal NLA saturates the bound of theoretical success probability, $P^{\mathrm{NLA}}_{\mathrm{succ}} \leq 1/g^2$. Our reasoning for this choice is that, so far, the NLA itself has not been experimentally demonstrated such that it is impossible to confidently make any realistic predictions as to its performance. By assuming it is ideal we can, for certain, provide theoretical, tangible limits pertaining to a protocol relying on its use. Of course, any physical NLA will undoubtedly fall short of this maximum success probability and the performance of QI with an NLA will degrade alongside it.

\begin{figure}[t!]
    \centering
x    \includegraphics[width=0.9\linewidth]{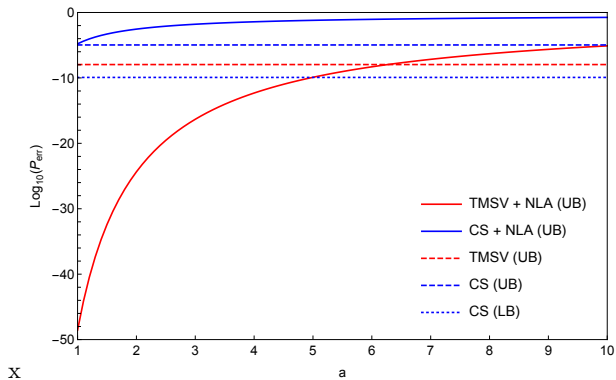}
    \caption{Error probability exponents for QI using a maximally-entangled TMSV source with a sub-optimal NLA (red, solid) at the receiver, compared to a coherent state source with the same NLA (blue, solid). We consider a non-ideal NLA whose probability of success is half the maximum, $P^{\mathrm{NLA}}_{\mathrm{succ}}=1/(a g^2)$, and results are plotted as function of the parameter $a$. Also included are associated upper bounds without the use of the NLA (dashed) showing how the NLA enhances this upper bound. The lower bound for coherent states (blue, dotted) shows the point beyond which a quantum enhancement (through the NLA, entanglement, or both) may be confirmed. As before, we have set $N_B=0.1$, $\kappa=0.2$ and are considering low-energy applications with $N_S=0.01$. All probabilities are evaluated at $M=10^4$.}
    \label{fig:suboptimalNLA}
\end{figure}

Fig.~\ref{fig:suboptimalNLA} plots the error exponents for the various protocols considered at a fixed number of uses, $M=10^4$. For the NLA protocols, with both TMSV and coherent state sources, the error exponent is plotted as a function of decreasing success probability. This is given in terms of the parameter $a$, defined via a reparameterisation of the success probability $P^{\mathrm{NLA}}_{\mathrm{succ}}(g,a) = 1/(a g^2)$, such that when $a=1$ we recover the ideal NLA performance given so far. The regime considered is the same as that studied previously in the low-energy limit: $N_S=1/100$, $N_B=1/10$ and $\kappa =1/5$. It can be seen that within this regime the advantages afforded by an NLA are robust with respect to its potential inefficiencies. The new quantum advantage that may be established here persists up to $a \simeq 5$. In other words, one can tolerate a reduction in efficiency up to $\sim 80$\% and still retain an advantage.

\section{Conclusion}

This paper has implemented the action of an NLA at the detection stage of the QI protocol for the purposes of quantum target detection. By mapping the resultant protocol to an equivalent one without the use of an NLA but modified effective parameters, the QCB for symmetric quantum hypothesis testing has been computed. This has been done for both the maximally-entangled TMSV state and the theoretically optimal classical benchmark, the coherent state, with comparisons made between the two assuming a theoretically maximal probability of success for the NLA.

Results show that the employment of non-Gaussian receivers for Gaussian sources in quantum target detection can be beneficial. In particular, an improvement in effective signal-to-noise ratio, resulting in a diminished error probability in target detection, occurs when the NLA is used with a TMSV source. Such results cannot be definitely achieved when the Gaussian source is the semi-classical coherent state; in this case, the performance is almost always bounded by the coherent state performance when no NLA is used which, in for applications in the optical domain, may be achieved through homodyne detection. There exist certain regimes where there is gain compared to the upper bound but, due to physical constraints imposed by our analysis, these do not hold with respect to the coherent state lower bound and thus cannot currently be confirmed.

The mapping used to compute the bounds results in a system of effective parameters for which a quantum advantage is not typically possible in a non-NLA protocol. At least, within such a regime the maximal advantage in error exponent certainly falls short of the potential value of 6 dB. In fact, within the regime considered here, comparing the upper bound to the lower bound of coherent state illumination which serves as our classical benchmark, it is possible to establish a $\sim 7$dB enhancement in the effective signal-to-noise ratio. Meanwhile, the equivalent non-NLA protocol here exhibits a $\sim 3$dB loss. Thus, the use of an NLA is able to not only amplify QI performances in regimes where the the potential gain is limited, but also establish new regimes of possible quantum advantage thereby extending the scope of of applicability of QI-based quantum target detection.

\textbf{Acknowledgments.}~This work has been funded by the European Union's Horizon
2020 Research and Innovation Action under grant
agreement No. 862644 (FET-Open project: Quantum readout techniques
and technologies, QUARTET). AK acknowledges sponsorship by EPSRC Award No. 1949572 and Leonardo UK. MG would like to acknowledge funding from the European Unions Horizon 2020 Research and Innovation Program under grant agreement No. 820466 (Continuous Variable Quantum Communications, `CiViQ’). The authors additionally thank Quntao Zhuang for their valuable feedback.

\appendix
\section{The quantum Chernoff bound (QCB)}\label{sec:QCBtools}

The binary decision between target absence and presence is
reduced to the discrimination of the two quantum states $\hat{\rho}_{R,I}^{i}$ with
$i=0,1$ \cite{cheflesQSD,barnettQSD,cheflesstrategies}. 

For Gaussian states, closed formulae exist for the computation of bounds on the minimum error probability in quantum state discrimination, such as the quantum Chernoff bound (QCB)~\cite{QCB}
\begin{align}
P_{\mathrm{min}}  &  \leq P_{\mathrm{QCB}}:=\frac{1}{2}\left(  \inf_{0\leq
s\leq1}C_{s}\right)^M  ,\nonumber\\
C_{s}  &  :=\Tr\left[  (\hat{\rho}_{R,I}^{0})^{s}(\hat{\rho}_{R,I}^{1}%
)^{1-s}\right]  , \label{QCB}%
\end{align}
where the minimisation of the $s$-overlap $C_{s}$ occurs over all $0\leq
s\leq1$, and we are considering the discrimination of $M$ mode conditional density operators. For the problem under study, the minimum is achieved for $s=1/2$ that
corresponds to the simpler quantum Bhattacharyya bound~\cite{RMP}
\begin{equation}
P_{\mathrm{QBB}}:=\frac{1}{2}\Tr\left[  \sqrt{\hat{\rho}_{R,I}^{0}}\sqrt
{\hat{\rho}_{R,I}^{1}}\right]^M  . \label{QBB}%
\end{equation}

Note that the QCB, and QBB, form exponentially tight upper bounds on the minimum error probability which converges in the limit $M\rightarrow \infty$. One can also consider the lower bound to the minimum error probability~\cite{fuchs1999cryptographic} which is, in general, not exponentially tight and takes the form
\begin{equation}
P_{\mathrm{min}} \geq \frac{1}{2} \left( 1- \sqrt{1- \Tr\left[  \sqrt{\hat{\rho}_{R,I}^{0}}\sqrt
{\hat{\rho}_{R,I}^{1}}\right]^{2M}}  \right). \label{errorLB}%
\end{equation}

Consider two arbitrary $N$-mode Gaussian states, $\hat{\rho}_{0}%
(\mathbf{x}_{0},\mathbf{V}_{0})$ and $\hat{\rho}_{1}(\mathbf{x}_{1}%
,\mathbf{V}_{1})$, with mean $\mathbf{x}_{i}$ and CM $\mathbf{V}_{i}$ with
quadratures $\mathbf{\hat{x}}=\left(  \hat{q}_{1},\hat{p}_{1},\dots,\hat
{q}_{N},\hat{p}_{N}\right)  ^{T}$ and associated symplectic form
\begin{equation}
\mathbf{\Omega}=\bigoplus_{k=1}^{N}%
\begin{pmatrix}
0 & 1\\
-1 & 0
\end{pmatrix}
.
\end{equation}
We can write the $s$-overlap as~\cite{pirandola2008computable}%
\begin{equation}
C_{s}=2^{N}\sqrt{\frac{\det\mathbf{\Pi}_{s}}{\det\mathbf{\Sigma}_{s}}}%
\exp\left(  -\frac{\mathbf{d}^{T}\mathbf{\Sigma}_{s}^{-1}\mathbf{d}}%
{2}\right)  ,
\end{equation}
where $\mathbf{d}=\mathbf{x}_{0}-\mathbf{x}_{1}$. Here $\mathbf{\Pi}_{s}$ and
$\mathbf{\Sigma}_{s}$ are defined as
\begin{equation}
\mathbf{\Pi}_{s}:=G_{s}(\mathbf{V}_{0}^{\oplus})G_{1-s}(\mathbf{V}_{1}%
^{\oplus}),
\end{equation}
\vspace{-0.5cm}
\begin{equation}
\mathbf{\Sigma}_{s}:=\mathbf{S}_{0}\left[  \Lambda_{s}\left(  \mathbf{V}%
_{0}^{\oplus}\right)  \right]  \mathbf{S}_{0}^{T}+\mathbf{S}_{1}\left[
\Lambda_{1-s}\left(  \mathbf{V}_{1}^{\oplus}\right)  \right]  \mathbf{S}%
_{1}^{T},
\end{equation}
introducing the two real functions
\begin{align}
G_{s}(x)  &  =\frac{2^s}{(x+1)^{s}-(x-1)^{s}}\nonumber\\
\Lambda_{s}(x)  &  =\frac{(x+1)^{s}+(x-1)^{s}}{(x+1)^{s}-(x-1)^{s}},
\end{align}
calculated over the Williamson forms $\mathbf{V}_{i}^{\oplus}%
:=\mathbf{\bigoplus}_{k=1}^{N}\nu_{i}^{k}\mathbf{1}_{2}$, where $\mathbf{V}%
_{i}^{\oplus}\mathbf{=S}_{i}\mathbf{\mathbf{V}}_{i}^{\oplus}\mathbf{S}_{i}%
^{T}$ for symplectic $\mathbf{S}_{i}$\ and $\nu_{i}^{k}\geq1$ are the
symplectic spectra~\cite{serafini2003symplectic,pirandola2009correlation}.
\end{document}